# Simulation of Open Quantum Dynamics with Bootstrap-Based Long Short-Term Memory Recurrent Neural Network

Kunni Lin[1], Jiawei Peng[2], Feng Long Gu[1,*] and Zhenggang Lan[2,*]

*1 Key Laboratory of Theoretical Chemistry of Environment, Ministry of Education, School of Chemistry, South China Normal University, Guangzhou 510006, P. R. China.*

*2 Guangdong Provincial Key Laboratory of Chemical Pollution and Environmental Safety and MOE Key Laboratory of Environmental Theoretical Chemistry, SCNU Environmental Research Institute, School of Environment, South China Normal University, Guangzhou 510006, P. R. China.*

*\* Corresponding Author. E-mail: gu@scnu.edu.cn; zhenggang.lan@m.scnu.edu.cn.*

**Abstract**

The recurrent neural network with the long short-term memory cell (LSTM-NN) is employed to simulate the long-time dynamics of open quantum system. The bootstrap method is applied in the LSTM-NN construction and prediction, which provides a Monte-Carlo estimation of forecasting confidence interval. Within this approach, a large number of LSTM-NNs are constructed by resampling time-series sequences that were obtained from the early-stage quantum evolution given by numerically-exact multilayer multiconfigurational time-dependent Hartree method. The built LSTM-NN ensemble is used for the reliable propagation of the long-time quantum dynamics and the simulated result is highly consistent with the exact evolution. The forecasting uncertainty that partially reflects the reliability of the LSTM-NN prediction is also given. This demonstrates the bootstrap-based LSTM-NN approach is a practical and powerful tool to propagate the long-time quantum dynamics of open systems with high accuracy and low computational cost.

**TOC graphics**

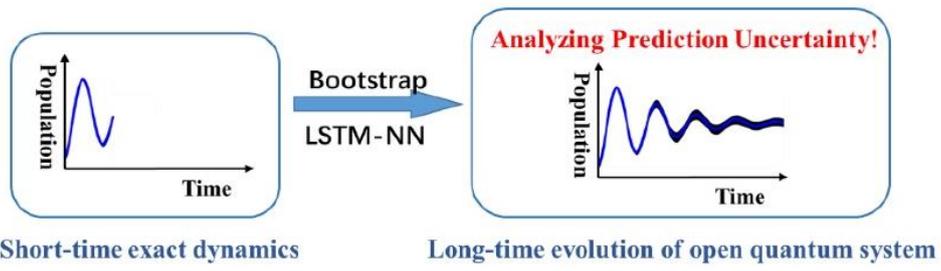

Quantum dissipative processes of complex systems widely exist in physics, chemistry and biology[1-5]. The theoretical descriptions of open quantum dynamics are extremely challenging[1-2], because a huge number of strong coupled degrees of freedom (DOFs) are involved. Within the system-plus-bath model, the quantum evolution of the reduced system under study is far more important than that of environmental DOFs, thus it is rather essential to develop a suitable dynamics approach for the correct description of the quantum evolution of open systems. Along this idea, various dynamics approaches were proposed in several decades[1-2, 6-29]. However, these available dynamics methods behave well in some situations, while they suffer from some disadvantages in other cases. For instance, some approximated approaches, such as mixed-quantum-classical Ehrenfest dynamics and its extension[6-7], and widely-used trajectory surface-hopping method[8-12], have rather low computational costs, while the inconsistent descriptions of the quantum and classical parts always bring some deficiencies, such as over-coherence problem[7, 13, 30]. The perturbative approaches, such as Redfield theory[1, 8], work well in weak system-bath coupling cases but fail in strong coupling situations. To obtain the precise description of the quantum evolution of open system, several numerically exact dynamics approaches were developed, including the hierarchical equations of motion (HEOM)[2, 14-16] and the hybrid stochastic-deterministic HEOM[29], the multiconfigurational time-dependent Hartree (MCTDH)[17] and its multilayer extension (ML-MCTDH)[18-21], tensor-network decomposition[22-24, 31], and so on[25, 32]. Although they are considered as rigorous theoretical approaches to simulate open quantum dynamics, these methods are generally limited by convergence problem or high computational cost. Therefore, it is still quite necessary to develop a theoretical approach that gives the reliable description of open quantum dynamics with high efficiency, without the restriction of model parameter regions.

In 2014, Gerrillo and Cao proposed the transfer tensor method[33] for the numerical analysis and processing of quantum dissipative dynamics. In the transfer tensor approach, a dynamical map was built to extract the key information of time-dependent dynamical correlation from the early-stage quantum evolution. All necessary dynamical information was compressed into transfer tensors, and these tensors can be used to propagate the reduced dynamics of the open system to long time scales under the assumption of the time-translational invariance. This pioneer work[33] figured out that the effective propagation of the quantum dynamics of open systems becomes possible using the dynamical map model built from the early-time evolution, if the early-stage dynamics already contains the enough information to describe the essential temporal features and the dynamical map model can capture such features. This transfer tensor approach received considerable research interests[34-37]. For instance, Geva, Cao and co-workers[37] built the transfer tensor models based on the short-time propagation of the mixed quantum-classical Liouville dynamics method, and used such models to propagate the long-time dynamics. This provided an accurate and robust approach to simulate the reduced dynamics of open quantum systems, by combining the accurate short-time propagation given by the mixed quantum-classical Liouville dynamics and the robust long-time propagation by the transfer tensor method.

As powerful techniques that can extract the essential features of input raw data and make the prediction of unknown outputs, machine learning algorithms[38-39] were widely applied in many scientific fields[39-50], such as physics, chemistry, computational science, bioinformatics and so on. Especially, the recurrent neural network (RNN)[39] displays the excellent ability to interpret the complex temporal behavior for time-series problems. By forming the neural-network structures with the feedback loop to accept both the input of the current step and the output of the previous step, the RNNs are able to store the historical information to process the prediction of future evolution. However, the unexpected vanishing and exploding gradient problems[51] limit the application of RNNs in long-time scale situations. To overcome such deficiency, the long short-term memory recurrent neural network (LSTM-NN)[39, 52] was proposed, which can model the long-range dependencies of the time-series data set, store this key information and make the forecasting for the future unseen data. Due to the excellent performance of LSTM-NNs, they have been widely used in various types of time-series analysis and forecasting problems[48-50], such as speech recognition and natural language processing. The applications of RNN and LSTM-NN in the simulation of the evolution of open quantum systems were conducted by Yang Zhao and co-workers[43-45]. This indicates that the LSTM-NN method may be an excellent approach to propagate the long-time dissipative dynamics of open quantum systems. In fact, the LSTM-NN approach can be viewed as a nonlinear map model, while the transfer tensor formulism is a linear map model. In principle, other nonlinear map models, such as the convolutional neural network (CNN)[39, 53] and kernel ridge regression (KRR)[54], may also be applicable in the propagation of quantum evolutions. In addition, we noticed that Banchi et. al.[46] once proposed to model the non-Markovian quantum processes by RNNs with Gated Recurrent Unit (GRU).

The application of the LSTM-NN method in the dynamics propagation of open quantum system certainly faces some challenges. One of the major limitations is that we do not have the clear idea about the prediction accuracy with time being, because the future reduced dynamics evolution is unknown. This largely prevents the wide application of the LSTM-NN method in the analysis and processing of the quantum dissipative dynamics. Thus, it is crucial to find a suitable way to evaluate the LSTM-NN forecasting quality as a function of time. The bootstrap resampling method[55-60] introduced by Efron et al. provides a Monte-Carlo approach to solve this key problem. An ensemble of fitting models was constructed under the help of the stochastic resampling based on the training data set. The averages of the predicting values by all trained models give the future data series and the distribution of them represents the confidence interval of prediction results. As a practical approach in the estimation of the prediction uncertainty, the bootstrap method gives a primary view to evaluate the forecasting reliability of fitting models, including LSTM-NNs.

In this work, we combine the LSTM-NN and bootstrap approaches for the accurate propagation of the quantum dissipative dynamics of reduced system to long-time scales. The key information of dynamical correlation features is extracted by LSTM-NNs in the basis of the short-time early-stage quantum evolution obtained from numerically exact ML-MCTDH method. When such essential dynamics information is captured and stored in the LSTM-NNs, the network models can be used to propagate the reduced dynamics of open quantum systems to a long-time scale. Most importantly, the bootstrap method is employed here. When a large number of LSTM-NNs are built with the resampling of the training data set, we not only obtain the prediction of the future dynamics but also acquire the forecasting uncertainty. This helps us to evaluate the reliability of the reduced dynamics propagation given by the LSTM-NNs. The performance of the bootstrap-based LSTM-NN is examined in several system-plus-bath models. By monitoring the prediction data set and forecasting uncertainty, it is possible to build a stable, accurate and flexible propagation of the long-time reduced dynamics of open quantum systems. This work demonstrates that the current protocol is a trustable and feasible approach to estimate the long-time quantum dissipative dynamics.

In this work, we considered to simulate the excited-state energy-transfer dynamics widely existing in solar energy conversion[1-2]. The system-plus-bath Hamiltonian was taken here to define the site-exciton model, which described two localized electronic states (system part) coupled with their individual vibrational modes (bath part). We employed the numerically exact ML-MCTDH method to simulate the quantum dissipative dynamics, which was performed by the Heidelberg MCTDH package[61]. More detailed discussions on Hamiltonian and ML-MCTDH were given in the Supporting Information (SI) and Figure S1 therein.

The initial condition was determined by vertical placing the lowest vibrational level of the ground state minimum to one localized excited electronic state. In the ML-MCTDH wavepacket propagation, the reduced density matrix of the electronic DOFs was calculated, in which the diagonal elements represent the electronic population and the off-diagonal elements describe the electronic coherence. We took the population difference $\Delta = \rho_{11} - \rho_{22}$, the real and imagery terms of off-diagonal elements [$\text{Re}\{\rho_{12}\}$ and $\text{Im}\{\rho_{12}\}$] to define the time-series vector-data set in the training and prediction processes.

The RNN is a class of artificial neural networks (ANNs)[39, 51], in which a directional connective graph takes outputs of the previous step as inputs of the current step. Because the memory information is taken into account, RNNs in principle can predict further events governed by the temporal features extracted from the history evolution.

The simplest RNN framework with rolled and unrolled configurations are displayed in Figure 1(a). At each time step $t$, the recurrent layer receives the inputs [$x_{(t)}$ and $h_{(t)}$] corresponding to their own outputs at the previous time step, and gives the output $y_{(t)}$. Therefore, the RNN model can be used to explore data pattern of a time series by the recurrence of

$$y_{(t)} = h_{(t)} = f(h_{(t-1)}, x_{(t)}). \tag{1}$$

However, due to the existence of vanishing and exploding gradient problems, the simple RNNs are not able to analyze the long-sequence time-series data. To solve this problem, a variation of RNN named LSTM-NN was developed, in which the so-called the LSTM cell is introduced by adding a few of control gates in each neuron[39, 51]. The architecture of this LSTM cell is shown in the Figure 1(b).

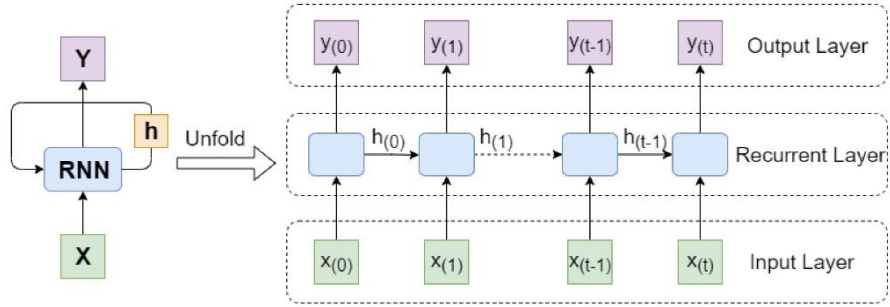

(a) The Simple RNN.

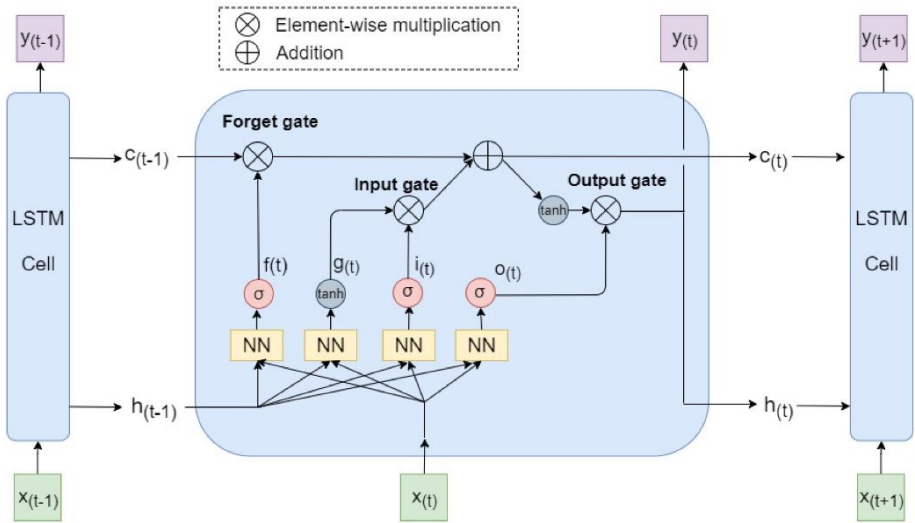

(b) The LSTM Cell.

Figure 1. (a) The simple RNN model rolled (left panel) and unrolled (right panel) configurations; (b) The LSTM cell.

The LSTM cell provides a more comprehensive description of the recurrence process. At time step $t$, the LSTM cell accepts the input vector $x_{(t)}$, as well as two additional vectors named as the short-term state vector $h_{(t-1)}$ and the long-term state vector $c_{(t-1)}$. After processing these vectors via three gates [Figure 1(b)], the LSTM cell generates the output vectors of $y_{(t)}$, $h_{(t)}$ and $c_{(t)}$. Next, both short-term and long-term state vectors become the inputs of the LSTM cell at the next step successively.

The mathematical details of the LSTM network are described by the following equations:

$$i_{(t)} = \sigma(W_{xi}^T x_{(t)} + W_{hi}^T h_{(t-1)} + b_i), \quad (2)$$

$$f_{(t)} = \sigma(W_{xf}^T x_{(t)} + W_{hf}^T h_{(t-1)} + b_f), \quad (3)$$

$$o_{(t)} = \sigma(W_{xo}^T x_{(t)} + W_{ho}^T h_{(t-1)} + b_o), \quad (4)$$

$$g_{(t)} = tanh(W_{xg}^T x_{(t)} + W_{hg}^T h_{(t-1)} + b_g), \quad (5)$$

$$c_{(t)} = f_{(t)} \cdot c_{(t-1)} + i_{(t)} \cdot g_{(t)}, \quad (6)$$

$$y_{(t)} = h_{(t)} = o_{(t)} \cdot tanh(c_{(t)}). \quad (7)$$

Here, the definition of variables $x_{(t)}$, $y_{(t)}$ and $h_{(t)}$ is given in Eq. 1. The LSTM cell includes three gates [Figure 1(b)], namely the input gate, the forget gate and the output gate. The input gate $i_{(t)}$ (Eq. 2) defines how much information of the inputs ($x_{(t)}$ and $h_{(t-1)}$) should be accepted in the current cell. The forget gate $f_{(t)}$ (Eq. 3) determines which part of input information should be thrown away or be remained. The output gate $o_{(t)}$ (Eq. 4) decides which information should be saved and transferred to the next cell. All three gates are controlled by the *sigmoid* activation function. At the same time, an intermediate vector $g_{(t)}$ (Eq. 5) is generated, which provides the additional controls of the data processing and information transfer via a *tanh* activation function. The tensors $W$ and $b$ are the weight and bias parameters of the LSTM cell, and different subscripts of these tensors are used to label their corresponding gate controls. The long-term state vector $c_{(t)}$ is updated according to Eq. 6. Eventually, the final output $y_{(t)}$ of current cell is obtained and the short-term state vector $h_{(t)}$ is given by using the output gate in Eq. 7. Several layers with LSTM cells are connected to give the final

LSTM-NN structure. All parameters in the LSTM-NN model can be determined by minimizing the deviation between the network output values and original data. In this way, the constructed LSTM-NN model allows us to capture the temporal pattern of the sequence data and to predict further events in the basis of historical dynamical feature.

As we realize, it is extremely crucial to estimate the confidence range of the model prediction, in order to access the reliability and stability in the forecasting of time-series data set. The prediction uncertainty is normally coming from various sources. One important error comes from the model uncertainty, which is relevant to the fact that model parameters in principle should follow a distribution. Thus, the parameter uncertainty must be taken into account to estimate such model uncertainty. This can be simply realized by the construction of the ensemble of proper parameter sets under the same LSTM-NN model structure. The other key error, the so-called model misspecification, is caused by the fact that the unseen prediction and training data sets may not follow the same distribution pattern. Such problem is normally remedied by the generation of different training subsets by resampling from the whole training data set, and this provides a Monte-Carlo description of different distribution data patterns.

As reported by previous works[62-64], the bootstrap theory is a powerful technique to generate uncertainty ranges of the machine learning models via the resampling of the training data set. In this approach, the random input sets are generated repeatedly by resampling over the initial training data set, and the size of each new sampled data set is the same as initial one. For example, consider a training data set as

$$X = [x_1, x_2, x_3, x_4, x_5, x_6, x_7, x_8], \qquad (8)$$

the bootstrap resampling approach may give us possible new training sampling sets as below:

$$X_1 = [x_1, x_3, x_3, x_2, x_2, x_2, x_7, x_8],$$

$$X_2 = [x_4, x_5, x_7, x_1, x_1, x_8, x_7, x_8],$$

$$X_3 = [x_6, x_2, x_4, x_4, x_5, x_7, x_7, x_8],$$

$$X_4 = [x_3, x_2, x_4, x_4, x_1, x_6, x_7, x_8],$$

………… (9)

In this way, the model misspecification is represented by the distribution of different input data sets generated by resampling. Since each set $(X_i)$ is used to train an independent model, the model parameters are re-fitted starting from the LSTM-NN with the same network structure. All of these machine learning models fitted from different training sets constitute a stochastic description of the model distribution. In this sense, the bootstrap method includes both the model uncertainty and model misspecification. Therefore, we may use such approach to estimate the confidence interval in the LSTM-NN forecasting of time-series data.

To obtain the long-time evolution of quantum dynamics, we need to build the LSTM-NN model from the short-time dynamics. At each step, the single data point corresponds to a vector with three components $[\Delta = \rho_{11} - \rho_{22}, \text{Re}\{\rho_{12}\}, \text{Im}\{\rho_{12}\}]$. The model prediction uncertainty is estimated by the bootstrap resampling approach. The detailed working procedure is given as below.

(a) *Data Set Preparation.*

First of all, the short-time quantum evolution was calculated with the numerically accurate ML-MCTDH method. Given a known temporal data series that represents the propagation of the reduced density matrix within the short-time ($T_t$) duration, we built a set of time series data sequences {$S_i$} with the length $L$ (see Figure 2), *i.e.,* each $S_i$ contained the $L$-number of successive data points in time order. In another word, the first element of $S_i$ was the *i*-th data point of the whole known temporal data series that represents the evolution within the time duration of $T_t$, and the last element is the (*i*+*L*-1)-th data point.

According to the chronological order, all sequences {$S_i$} were divided into two parts in a ratio of 3:1, in which the first part (Group A in Figure 2) with early time order was employed to build the LSTM-NN model, while the second part (Group B in Figure 2) with later time order was used to evaluate the model performance. All sequences in the first part (Group A) were randomly shuffled, while the element order in each sequence remains unchanged to keep the temporal order information. Next all sequences in this set were divided into two groups in a ratio of 7:3, giving the training set (**T**$_{set}$) and the internal validation set (**V1**$_{set}$), respectively. All sequences of the second part (Group B) with the later time order give us the external validation set (**V2**$_{set}$). Two different validation sets (**V1**$_{set}$ and **V2**$_{set}$) were considered in the current work and their essential roles are discussed in below. We provide a brief scheme of the time-series division in Figure 2.

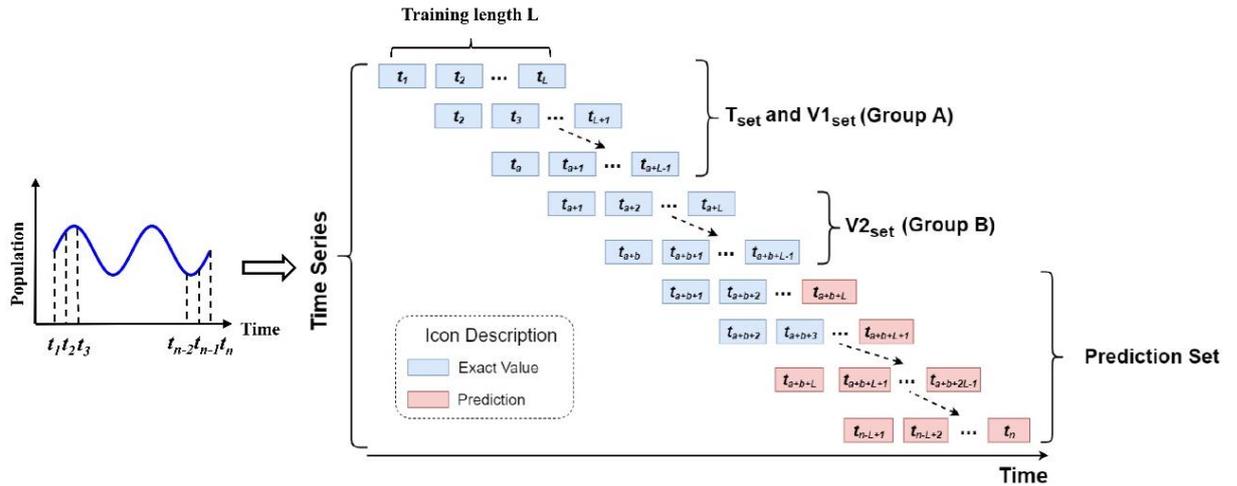

Figure 2. A brief scheme of the time-series division.

(b) *Primary Model Construction and Selection.*

The training set **T**$_{set}$ and the first validation set **V1**$_{set}$ were used to fit LSTM-NN models. The training of the LSTM-NN was performed in the basis of the training set **T**$_{set}$, in which the mean square error (MSE) loss function and the adaptive moment estimation (Adam) optimizer were used. To avoid the overfitting, the early stop approach was used according to the first validation set **V1**$_{set}$.

Several LSTM-NN models with different network structures were built initially by varying the numbers of layers and neurons. After all models were built in the training step, their performances were examined based on the second validation set **V2**$_{set}$. This allows us to select a few of LSTM-NN models with very small validation errors. Although these models may display different structures, all of them show the reasonable prediction of the data series for **V2**$_{set}$. Importantly, it is necessary to emphasize that each proper LSTM-NN model is associated with its own optimal length $L$. Thus, the optimal choices of the length $L$ of the data set sequence $\{S_i\}$ were also examined. Essentially, all above training and validation procedure can be viewed as a "grid searching" approach in the LSTM-NN model construction for the optimal choices of the network topology (layer and neuron numbers) and the length $L$ of the data sequence. More details are given in SI.

The construction and prediction of the LSTM-NN model were conducted by our home-made Python code base on TensorFlow[65] and Keras[51, 66]. In LSTM-NN training step, the number of epochs were 300 with the batch size of 50, and this setup display the optimal performances in our trial calculations. The nonlinear activation $tanh$ function was used on each hidden layer. The LSTM-NN model parameters (the number of layers and neurons) and the optimal length $L$ of the time sequences $\{S_i\}$ were selected by the grid search in the basis of **V2$_{set}$**. In this way, 20 LSTM-NN models were selected, which display the lowest validation error. Then the error distribution of the chosen LSTM-NN models were plotted and their average deviation was calculated. Finally, only the LSTM-NN models with the error bellow this average deviation was taken, because in principle each model with small error should represent a function close to the target function in fitting theory. In this sense, we also considered another dimension of the model uncertainty by including different neural network structures in the final model prediction.

(c) *Prediction of Future Evolution with Bootstrap*.

Starting from each chosen LSTM-NN model and its corresponding optimal length $L$ of each training date sequence, an ensemble of LSTM-NN models were built using the bootstrap resampling method. In this step, all sequences of the training set (**T$_{set}$**) and the internal validation set (**V1**$_{set}$) were mixed again to give the original set (**BT**$_{set}$) of data sequences for bootstrap resampling. Based on each resampled data set, an individual LSTM-NN network was re-fitted under the same network structure as the original one. In this way, an ensemble of 100 independent LSTM-NNs were obtained by the bootstrap for each chosen LSTM-NN.

After the bootstrap approach was employed in every chosen LSTM-NN model given by Step (b), several groups of LSTM-NNs were generated at the end. All of them were employed to simulate the unseen future time-series data sequence in the long-time quantum dynamics. The average over all LSTM-NN predictions gave the further evolution of temporal data. At the same, the confidence interval of the model prediction was estimated by the standard deviation of forecasting values given by all LSTM-NN predictions. In this sense, the LSTM-NN model with the bootstrap method not only allows us to generate the unseen quantum dynamics evolution up to a long- time scale, but also gives a crucial approach to measure the forecasting confidence in statistical manner.

We first examine the performance of the LSTM-NNs combined with bootstrap resampling approach in the simulation of open quantum dynamics. The first model is composed of two electronic states with unbiased energy levels and the interstate coupling of $V_{12}$=0.0124 eV. The Debye-type spectral density with the characteristic frequency $\omega_c$=200 cm$^{-1}$ and the reorganization energy with $\lambda$=25 cm$^{-1}$ are taken to characterize electron-phonon couplings. Here each electronic state couples with its own individual bath represented by 100 discrete modes and totally 200 modes were included in the site-exciton model. The above parameters in the bath mode discretion should be enough for the current purpose, because they are only used to generate a model for the quantum evolution, instead of giving the precise description of the full dynamics in the continuous bath limit.

First, we stopped the ML-MCTDH quantum evolution at 350 fs and employed the electronic density matrix to build the bootstrap-based LSTM-NNs. As shown in Figure 3(a) and Figure S2 in SI, the LSTM-NN result gives very small fitting error within the training and internal validation time duration. The minor error appears within 250-350 fs (Figure S2 in SI), because this period corresponds to the external validation set that is not included in the training step. For the long-time propagation, the quantum dynamics modelled by the bootstrap-based LSTM-NNs is roughly consistent with those obtained in the exact quantum evolution by ML-MCTDH, see Figure 3(a), while their difference becomes noticeable with time being. In addition, the confidence interval increases obviously, and after 500 fs the prediction uncertainty becomes quite large. Overall, the LSTM-NN simulation of the quantum evolution is reasonable within the very short-time scale, while the result becomes not reliable for the long-time dynamics due to the increasing of prediction uncertainty with time being.

Next, we took the longer-time propagation dynamics up to 450 fs to build the LSTM-NNs. The fitting and prediction results, along with the full quantum dynamics, are displayed in Figure 3(b) and Figure S2 in SI. The training errors are extremely small, and we also noticed that validation errors calculated in the basis of the external validation set remain negligible. In the current case, the long-time quantum dynamics by LSTM-NNs is quite accurate even up to 1 ps, with respect to the full quantum dynamics evolution, as shown in Figure 3(b). At the same time, the prediction uncertainties for all forecasting elements of reduced electronic density matrix are rather minor, much smaller than those based on shorter-time (350 fs) dynamics, see Figure S2 in SI.

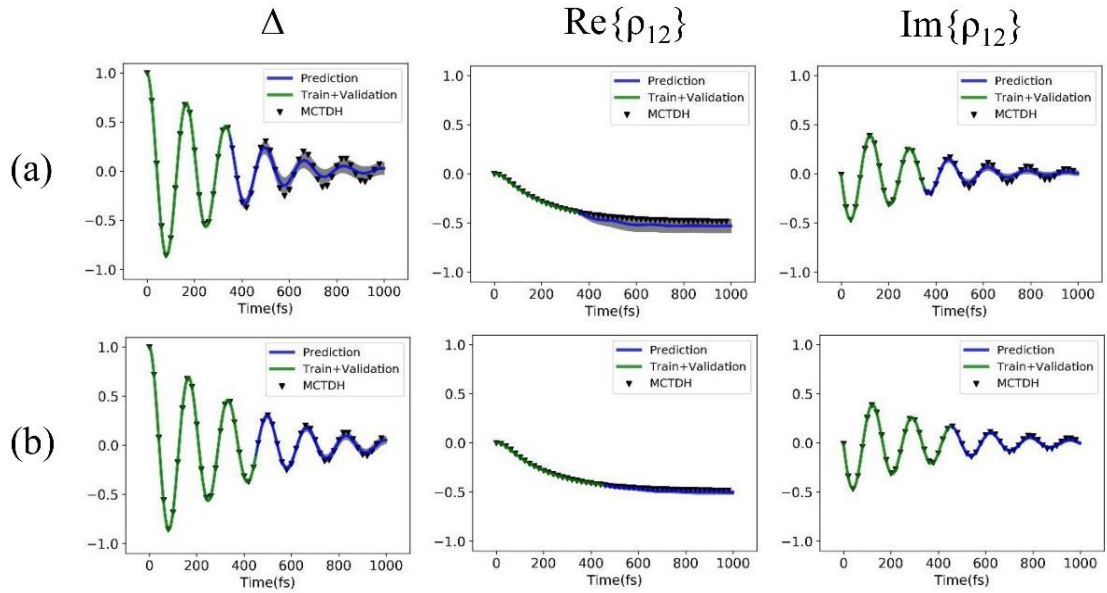

Figure 3. The quantum dynamics modelled by the bootstrap-based LSTM-NNs *vs.* the ML-MCTDH quantum dynamics with the model of $V_{11}-V_{22}=0$ eV, $V_{12}=0.0124$ eV, $\omega_c=200$ cm$^{-1}$ and $\lambda=25$ cm$^{-1}$. Different training/validation data lengths are shown, namely 350 fs (a) and 450 fs(b). The green lines denote the training and validation samples used in the LSTM-NN construction. The blue lines correspond to the LSTM-NN modelling of the future dynamics. The black triangles display the ML-MCTDH simulation results. And the grey region represents the confidence interval of the forecasting results.

However, caution should be paid in the above comparison in Figure 3. For instance, when we took the dynamics data with the length of 350 fs or 450 fs to build the LSTM-NNs, it is not fair to examine the performance of these two sets of LSTM-NN models by checking both of their prediction results up to 1 ps. Since we assume that the time duration of known dynamics is $T_t$, it may be more reasonable to check the accuracy and uncertainty of the LSTM-NN dynamics at the time $1.5T_t$, $2.0T_t$ and so on. However, in this comparison set, the longer training length (450 fs) still provides much better LSTM-NN models to simulate the long-time dynamics than the shorter training length (350 fs), as shown in Figure S3 in SI.

As discussed above, the length of the time-series data from the known early-stage dynamics largely influences the prediction ability of the LSTM-NNs. When the time duration of the early-stage quantum propagation is too short to represent the underline dynamical correlation, it is not possible to build an accurate dynamical model based on such short-time dynamics. As the consequence, the simulation results of the unknown future dynamical feature are not trustable, and at the same time the large forecasting uncertainty occurs. Only when the historical data set used in the model construction is longer enough to capture the underline dynamical correlation, the time-dependent feature of the quantum dynamics can be well extracted by the LSTM-NNs. For instance, for the given example in Figure 3(b), the early-stage data set up to 450 fs contains the enough information to characterize the underlining quantum dynamics features. The reliable models can be constructed, which give the correction description of the long-time quantum evolution, as well as the small prediction error.

In principle the prediction reliability should decreases with time being, and sooner or later the forecasting results may become trustless, no matter which NN model is constructed. Thus, the employment of the RNN or LSTM-NN in the prediction of a time-series data set always suffers from the unknown accuracy in practices. In current work, the employment of the bootstrap resampling method partially alleviates this problem, because we may easily check the reliability of the forecasting models by monitoring the prediction uncertainty with time being. If the LSTM-NNs are built based on the early-stage quantum dynamics, it is always possible to examine the confidence interval for the prediction of the unknown future evolution. At least, when such uncertainty becomes large enough at a particular time, we immediately realize that the simulation result is not trustable anymore after such time step. On the other hand, if we wish to get the more-or-less accurate dynamics feature up to a given time step (such as 1 ps), it is necessary to check how long of the time-series data set ($T_t$) should be taken in the LSTM-NN construction. This can be realized by examining the forecasting uncertainty, after the LSTM-NNs are built based on different $T_t$.

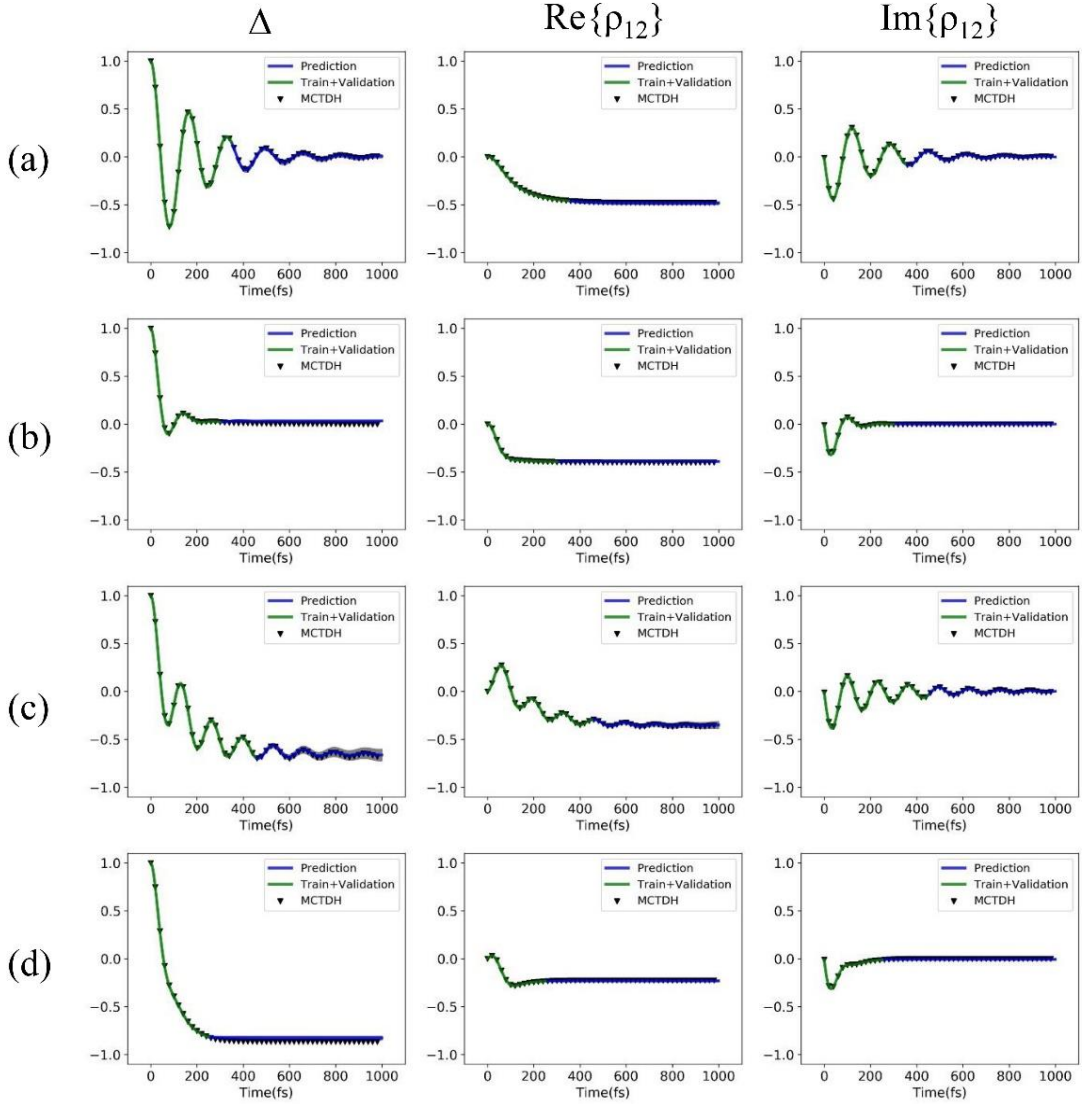

Figure 4. The quantum dynamics predicted by the bootstrap-based LSTM-NNs *vs.* the ML-MCTDH quantum dynamics with the model of (a) $V_{11}-V_{22}$=0 eV, $V_{12}$=0.0124 eV, $\omega_c$=200 cm$^{-1}$ and $\lambda$=49 cm$^{-1}$; (b) $V_{11}-V_{22}$=0 eV, $V_{12}$=0.0124 eV, $\omega_c$=200 cm$^{-1}$ and $\lambda$=196 cm$^{-1}$; (c) $V_{11}-V_{22}$=0.0186 eV, $V_{12}$=0.0124 eV, $\omega_c$=200 cm$^{-1}$ and $\lambda$=49 cm$^{-1}$; (d) $V_{11}-V_{22}$=0.0186 eV, $V_{12}$=0.0124 eV, $\omega_c$=200 cm$^{-1}$ and $\lambda$=225 cm$^{-1}$. The green lines represent the training samples and the validation samples. The blue lines correspond to the prediction results used LSTM-NN. The black triangles display the exact simulation dynamics trajectories calculated by ML-MCTDH. And the grey region represents the confidence interval of the prediction results at the current time.

The above results clearly indicate that the bootstrap-based LSTM-NN method provides a practical and reliable approach to simulate the quantum evolution to a long-time scale. This idea was also confirmed by the simulation of the quantum evolution in several models with different parameters. These models cover different state energy differences, interstate coupling strengths and electron-phonon couplings. Following the procedure described above, the proper bootstrap-based LSTM-NN construction is realized, which gives the excellent results for the long-time quantum dynamics of all models under study, see Figure 4. When the suitable ways to build the LSTM-NN are taken, the forecasting uncertainty always remains very small, which gives us the confidence in prediction results. This is further confirmed by the small deviation between the quantum evolution given by LSTM-NNs and ML-MCTDH methods.

In many realistic applications, it is not easy to obtain the exact numerical results of the time-dependent dynamics propagation in the asymptotic limit. In these situations, the density matrix of the reduced system at the long-time limit becomes exactly the thermally-equilibrium reduced density matrix, which can be obtained using imaginary-time path-integral simulations[67]. This asymptotic limit may provide an independent and stringent test for the performance of the current bootstrap-based LSTM-NN approach. Certainly, the non-Markovian dynamics of open quantum systems with very low temperature and strong system-bath coupling is an extremely challenging topic in quantum dissipative dynamics. The hybrid stochastic-deterministic method is a promising approach to simulate such type of non-Markovian dynamical evolution, which combines the stochastic and HEOM approaches[29]. To give the preliminary evaluation of the performance of our proposed bootstrap-based LSTM-NN method, we also tried to simulate the non-Markovian dynamics with low temperature and strong system-bath coupling. As shown in Figure S4 in SI, the results look promising and the LSTM-NN approach can also give the reasonable prediction on the long-time dynamics, as well as the limit given by imaginary-time path-integral simulations[67].

The formalism of the transfer tensor method depends on the initial condition, *i.e.* the correlated and direct-product system-bath initial states lead to different dynamical evolutions. Certainly, the application of the bootstrap-based LSTM-NN approach displays the similar features. When different initial conditions are given, the suitable LSTM-NNs should be re-constructed and the bootstrap procedure should be re-performed to obtain the forecasting uncertainty.

In this work, we design the bootstrap-based LSTM-NN approach to simulate the long-time reduced dynamics of open quantum system. The LSTM-NNs capture the key dynamical correlation features from the short-time early-stage quantum evolution obtained with numerically exact ML-MCTDH method. As all essential dynamical features are stored in the LSTM-NNs, the network can be used for the accurate simulation of the reduced dynamics to a rather long-time scale. To clarify the prediction reliability of LSTM-NNs, the bootstrap method is employed. After the generation of a huge number of LSTM-NNs under the help of resampling training data, the bootstrap-based LSTM-NN approach provides both the propagation of the future dynamics and the forecasting uncertainty. This provides us a unique way to judge the reliability of the forecasting reduced dynamics.

The performance and feasibility of the bootstrap-based LSTM-NNs are benchmarked in several system-plus-bath models. For all model under study, it is possible to construct the proper LSTM-NN models for the reliable propagation of quantum evolution up to a rather long-time scale. We notice that the stable, accurate and flexible propagation of the long-time reduced dynamics of open quantum system can be achieved by monitoring the prediction data set and forecasting uncertainty. This work demonstrates the current bootstrap-based LSTM-NN protocol is a trustable, feasible and robust approach to estimate the long-time reduced dynamics of open quantum systems.

**Supporting Information Available:** Several relevant information: the brief explanation of the Hamiltonian and ML-MCTDH method; the details of the grid search in the model construction; the predicted value distribution calculated with the bootstrap method; the prediction confidence interval evoluting with time being; the prediction of the non-markovian dynamics with low temperature and strong system-bath coupling.

**Author Information**

**Corresponding Author**

E-mail: gu@scnu.edu.cn; zhenggang.lan@m.scnu.edu.cn.

**Notes**

The authors declare no competing financial interest.

**Acknowledgments**

The authors express sincere to thank the National Natural Science Foundation of China (No. 21873112, 21933011 and 21673085), the National Key Research and Development Program of China (2017YFB0203403), Guangdong Province Universities, and Colleges Pearl River Scholar for financial support.

*Rev. B* **2012,** *85* (11), 115412.

# Supporting Information for

# Simulation of Open Quantum Dynamics with Bootstrap-Based Long Short-Term Memory Recurrent Neural Network

Kunni Lin[1], Jiawei Peng[2], Feng Long Gu[1,*] and Zhenggang Lan[2,*]

*1 Key Laboratory of Theoretical Chemistry of Environment, Ministry of Education; School of Chemistry, South China Normal University, Guangzhou 510006, P. R. China.*

*2 Guangdong Provincial Key Laboratory of Chemical Pollution and Environmental Safety and MOE Key Laboratory of Environmental Theoretical Chemistry, SCNU Environmental Research Institute, School of Environment, South China Normal University, Guangzhou 510006, P. R. China.*

*\* Corresponding Author. E-mail: gu@scnu.edu.cn; zhenggang.lan@m.scnu.edu.cn.*

**1. Hamiltonian.**

In current work, the system-plus-bath Hamiltonian was taken in the description of the site-exciton model, in which the system part included two local-excited (LE) electronic states and the bath is composed of vibrational modes

$$H = H_S + H_B + H_{SB}. \quad (1)$$

The electronic part is written as

$$H_S = \sum_{k=1}^{2} |\varphi_k\rangle V_{kk} \langle\varphi_k| + \sum_{k \neq l} |\varphi_k\rangle V_{kl} \langle\varphi_l| \quad (2)$$

where $V_{kk}$ represents the energy of the excited state and $V_{kl}$ represents the electronic coupling of them. We assumed that two LE electronic states couple with their own individual bath modes. The harmonic oscillator bath is written as below

$$H_B = \sum_{k=1}^{2} \sum_{j}^{N_b} \frac{1}{2} \omega_{kj} (Q_{kj}^2 + P_{kj}^2), \tag{3}$$

and the bilinear electron-phonon interaction is considered as

$$H_{SB} = \sum_{k=1}^{2} |\varphi_k\rangle \left( \sum_{j}^{N_b} k_{kj} Q_{kj} \right) \langle \varphi_k|. \tag{4}$$

In the above two equations (Eq.3 and Eq. 4), the total number of bath modes is given by $N_b$ for each electronic state. Three parameters, $\omega_{kj}$, $Q_{kj}$, and $P_{kj}$, denote the corresponding frequency, position and momentum of each bath mode, respectively. The $k_{kj}$ characterizes electron-phonon coupling strength. The subscripts $k$ and $j$ refer to the indices of the electronic state and the bath mode, respectively.

The bath is characterized by the Debye-type spectral density:

$$J(\omega) = \frac{2\lambda \omega \omega_c}{\omega^2 + \omega_c^2}, \tag{5}$$

where $w_c$ and $\lambda$ refer to the characteristic frequency and the reorganization energy, respectively. It is possible to represent the spectral density by a series of discretized bath modes, namely

$$J_k(\omega) = \frac{1}{2} \pi \sum_{i=1}^{N} k_{ki}^2 \delta(\omega - \omega_{ki}). \tag{6}$$

Therefore, when the sampling interval $\Delta \omega$ is given, the coupling strength of each mode $k_{ki}$ is evaluated by the following equation:

$$k_{ki} = \left( \frac{2}{\pi} J_k(\omega_{ki}) \Delta \omega \right)^{1/2}. \tag{7}$$

In all models, we took the frequency domain as 0-1200 cm⁻¹ and $\Delta \omega$=12 cm⁻¹ to build the discreted bath modes.

## 2. ML-MCTDH.

As a numerical exact approach to treat the quantum dynamics of high-dimensional systems, the MCTDH approach and its extension ML-MCTDH can be viewed as the tree tensor decomposition in the representation of the total wavefunction, which largely save the computational cost in quantum dynamics propagation.

In the MCTDH framework, the time-dependent high-dimensional wavefunction is expanded as

$$\psi(Q_1,\cdots,Q_f,t) = \sum_{j_1=1}^{n_1}\cdots\sum_{j_f=1}^{n_f} A_{j_1,\cdots,j_f}(t) \sum_{k=1}^{f} \varphi_{j_k}^{(k)}(Q_k,t), \qquad (8)$$

where the degrees of freedom (DOFs) are represented by $Q_1,\cdots,Q_f$. Besides, $A_{j_1,\cdots,j_f}$ and $\varphi_{j_k}^{(k)}$ are the time-dependent expansion coefficients and the time-dependent basis functions, respectively. As the powerful extension of MCTDH, the ML-MCTDH method further expands the time-dependent basis functions recursively as

$$\varphi_m^{l-1;k_1\cdots k_{l-1}}\left(Q_{k_{l-1}}^{l-1;k_1\cdots k_{l-2}},t\right) = \sum_{j_1=1}^{n_1}\cdots\sum_{j_{p_{kl}}=1}^{n_{kl}} A_{m;j_1\cdots j_{p_{kl}}}^{l;k_1\cdots k_{l-1}}(t) \prod_{k_l=1}^{p_{kl}} \varphi_{j_{kl}}^{l;k_1\cdots k_l}\left(Q_{k_l}^{l;k_1\cdots k_{l-1}},t\right) \quad (9)$$

where $l$ represents the number of expanded layers; $k_1,\cdots,k_{l-1}$ connecting each node on the top layer and the particular primary coordinate denote the indices of the logical DOF. In the ML-MCTDH framework, the wavefunction is finally represented by a multilayer tree structure, as shown in Figure S1.

Although the ML-MCTDH is a powerful approach in the numerically exact treatment of the quantum dynamics of high-dimensional systems, it is rather challenging to use this approach to obtain the reasonable results of quantum dynamics. In practices, the convergence test must be performed by modifying the tree structure and the basis number in each node. In addition, the great attention should be paid to build the ML-MCTDH tree expansion, because different tree expansions may result in the significant difference of the computational costs, as well as distinguishing numerical convergence behaviors.

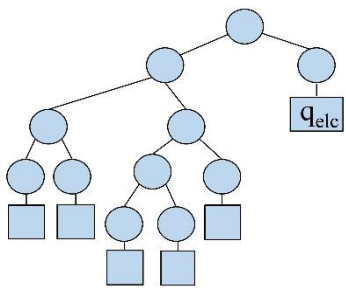

Figure S1. The ML-MCTDH tree structure in the expandion of wavefunction.

## 2. The Grid Search.

In the LSTM-NN model construction, the reasonable model structures and the training length $L$ of the data set sequence were obtained by the grid-search method, which is a common approach to find optimal neural network (NN) models. The details were given as below.

We considered three parameters, including the number of NN layers, the number of neurons in each layer and the length $L$ for the time-series data. The selection of these parameters was based on the below rules.

- For a given data sequence with the length $L$, the input of the LSTM-NN network is the data sequence with the length of $L$-1, while the output gives the last data point of this sequence. This automatically defines the setups of the input and output layers. During the model training process, the input-data shape of the input layer is [Batch Size * $L$-1 * Feature Dimension of Input Data], and the output-data shape of the output layer is [Batch Size * 1 * Feature Dimension of Input Data].

- The number of hidden layers was chosen as 1, 2 or 3.

- Each hidden layer has the same number of neurons. The maximum and minimum numbers of neurons for each layer were set between 110 and 10, and the spacing is 20.

- In the description the time-dependent dynamics, the whole time series was built by selecting snapshots at every 0.5 fs, which contains the quantum evolution within the early-time duration of $T_t$. The length of the known time series is given by $M=T_t/0.5$. The length $L$ of each individual sequence of time-series data was defined as follows. The minimum length $L$ corresponded to the time duration of 25 fs, and the maximum length was less than the half of the whole time series used in the training step. This defines $25/0.5 \leq L \leq M/2$ and the space of the $L$ value ($\Delta L$) was chosen as 25 fs.

By using the above grid search method, it is able to determine all parameters in the LSTM-NN construction. Although a large number of calculations were required, such process is very efficient, because all calculations are independent.

## 3. Time-Dependent Confidence Interval.

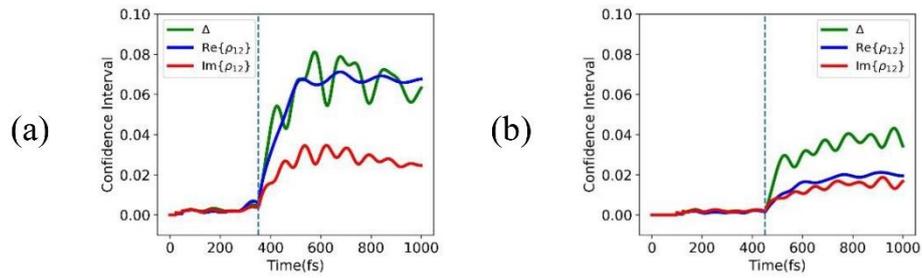

Figure S2. The preditction uncertainty calculated by boostrap-based LSTM-NN method, in which the RNNs were built based on the time-series data set with different lengths: (a) 350 fs and (b) 450fs. The vertical dashed lines denote the division between the trainning/validations and unknown data sets. The left and right panels correspond to the two subfigures in Figure 3 in the main manuscript.

## 4. The Result of the Prediction with Bootstrap.

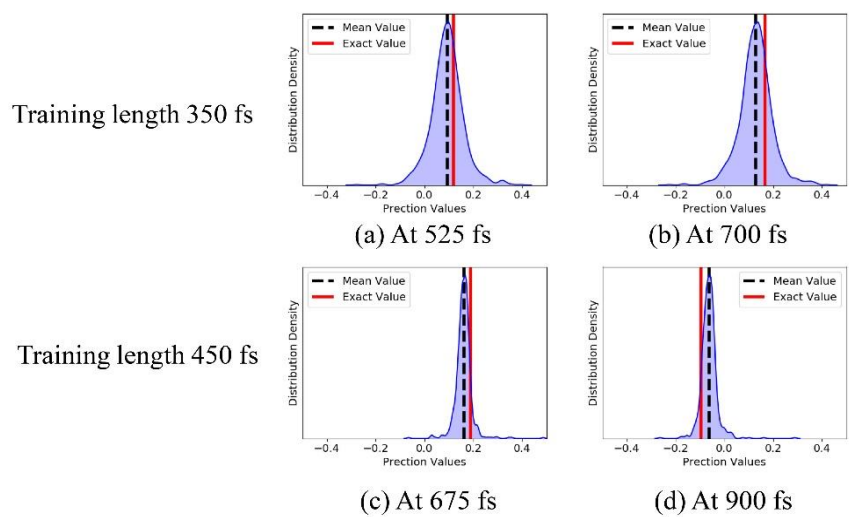

Figure S3. The examples of the predictions distribution given by bootstrap method at (a) 525 fs and (b) 700 fs with 350 fs training length, (c) 675 fs and (d) 900 fs with 450 fs training length. The black dashed corresponds to the meaning values of all prediction results. The red dashaed represents the label value given by the ML-MCTDH method. The upper and lower panels correspond to the two subfigures in Figure 3 in the main manuscript.

# 5. The Prediction of the Non-Markovian Dynamics with Low Temperature and Strong System-Bath Coupling.

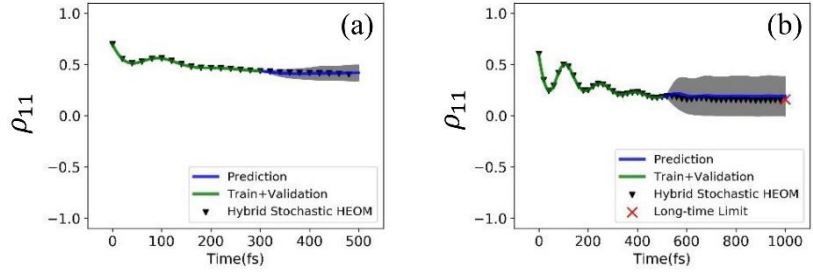

Figure S4. The quantum dynamics of the spin-boson model with $E$=50 cm$^{-1}$, $J$=100 cm$^{-1}$, $\omega_c$=53 cm$^{-1}$ and $\lambda$=100 cm$^{-1}$ by the bootstrap-based LSTM-NNs *vs.* the hybrid stochastic-deterministic HEOM quantum dynamics: (a) T= 300 K; (b) T=10 K. The green lines represent the training samples and the validation samples. The blue lines correspond to the prediction results used LSTM-NN. The black triangles display the exact simulation dynamics trajectories calculated by hybrid stochastic-deterministic HEOM method in reference[1]. The grey region represents the confidence interval of the prediction results at the current time. And the red fork in (b) corresponds to the long-time limit given by imaginary-time path-integral simulations.

The theoretical description of the non-Markovian dynamics with very low temperature and strong system-bath coupling represents the great challenging. To provide the preliminary view on the performance of the bootstrap-based LSTM-NN approach in this situation, we try to conduct the simulation based on the data of non-Markovian dynamics obtained by the hybrid stochastic-deterministic HEOM method[1]. The spin-boson model is taken here, in which the system Hamiltonian is $H_S = E\sigma_z + J\sigma_x$ and the same type spectral density of the bath is employed.

The off-diagonal element of the density matrix was not shown in the reference[1]. As the short-time population evolution cannot provide all information of the dynamical correlation, all dynamical maps constructed from such short-time evolution miss some dynamical information. In this case, the below tricks are taken to improve the performance of the LSTM-NN models.

- The longer-time duration is taken in the model training/validation step.
- A three-dimensional vector $\rho_{11}$, $\rho_{22}$, $\rho_{11}$-$\rho_{22}$ is used in the model construction, instead of only one element.
- In this work, a few of LSTM-NN models with small validation errors are chosen based on the second validation set, and these chosen LSTM-NN models are taken to perform the bootstrap task. Here, the additional LSTM-NN model selection step is conducted. After the bootstrap, an ensemble of the NNs with the fixed topology defines an uncertainty of this NN ensemble, and thus different NN ensembles display different uncertainties. We only collected the ensembles with the rather low uncertainty to estimate the overall confidence interval.

Although the current simulation suffers from the missing of the off-diagonal elements of density matrix, the long-time dynamics predicted by the LSTM-NN still looks promising, as seen in Figure S4 (a) and (b). For the low-temperature situation, the average of the bootstrap based LSTM-NN models gives the satisfied result in the population dynamics, which also agrees well with the long-time limit obtained by the imaginary path integral[2]. However, the current results are highly approximated. It is clear that the quantum coherence is important in this situation, which is characterized by the off-diagonal elements of the density matrix. The visible prediction uncertainty in Figure S4(b) may highly be relevant to the missing of this term in the LSTM-NN model construction.